\documentclass[10pt,a4paper,twoside]{article}

\usepackage{indentfirst}			
\usepackage{graphicx}				
\usepackage{amsgen,amsfonts,amssymb,amsbsy}	

\newenvironment{resum}{\begin{quote}\small}{\end{quote}}

\newcommand{\bfsf}[1]{\textsf{\textbf{#1}}}

\begin{document}

\thispagestyle{plain}		

\begin{center}


{\LARGE\bfsf{The Line Elements in the Hubble Expansion}}

\bigskip


\textbf{Moshe Carmeli}$^1$


$^1$\textsl{Department of Physics, Ben Gurion University, Beer Sheva 84105, 
Israel\\
(Email: carmelim.bgumail.bgu.ac.il)}
\\

\end{center}

\medskip


\begin{resum}
In this lecture I present the line elements that express the Hubble 
expansion. The coordinates here are the spatial coordinates $x$, $y$, $z$,
and the velocity coordinate $v$, which are actually what astronomers use in
their measurements. Two such line elements are presented: the first is the 
empty space (no matter exists), and the second with matter filling up the
Universe. These line elements are the comparable to  the standard ones, the
Minkowski and the FRW line elements in ordinary general relativity theory. 
\end{resum}

\bigskip


\section{Introduction}

In the standard cosmological theory one uses the Einstein concepts of space 
and time as were originally introduced for the special theory of relativity
and later in the general relativity theory. According to this approach all 
physical quantities are described in terms of the continuum spatial 
coordinates and the 
time. Using the general relativity theory a great progress has been made in
understanding the evolution of the Universe. 

Accordingly in the standard cosmological model one has the 
Friedmann-Robertson-Walker 
(FRW) line element
\begin{equation}
ds^2=dt^2-a^2\left(t\right)\left\{\frac{dr^2}{1-kr^2}+r^2\left(d\theta^2+
\sin^2\theta d\phi^2\right)\right\},
\label{eq.01}
\end{equation}
where $k$ is a constant,
$k<0$, open Universe,
$k>0$, closed Universe, 
$k=0$, flat Universe.
Here $a(t)$ is a scale function, and by the Einstein field equations, it satisfies
\begin{equation}
\left(\frac{\dot{a}}{a}\right)^2=\frac{8\pi}{3}G\rho-\frac{k}{a^2},
\label{eq.02}
\end{equation}
\begin{equation}
\ddot{a}=-\frac{4\pi}{3}G\left(\rho+3p\right)a,
\label{eq.03}
\end{equation} 
\begin{equation}
\frac{d}{dt}\left(a^3\rho\right)=-p\frac{d}{dt}\left(a^3\right),
\label{eq.04}
\end{equation}
where $\rho$ is the mass density, $p$ is the pressure, $G$ is Newton's 
constant, and the overdot denotes derivative with respect to $t$. Notice 
that the above equations are not independent. 

It is well known that both Einstein's theories are based on the fact that light
propagates at a constant velocity. However, the Universe also expands at a
constant rate when gravity is negligible, but this fact is not taken into 
account in Einstein's theories. Moreover, cosmologists usually measure spatial 
distances 
and redshifts of 
faraway galaxies as expressed by the Hubble expansion. In recent years this
fact was undertaken to develop new theories in terms of distances and velocities
(redshift). While in Einstein's special relativity the propagation of light  
plays the major role, in the new theory it is the expansion of the Universe 
that takes that role. It is the concept of cosmic time that becomes crucial in 
these recent theories. In the standard theory the cosmic time is considered to 
be absolute. Thus we talk about the Big Bang time with respect to us here on
Earth as an absolute quantity. Consider, for example, another galaxy that 
has,
let us say, a relative cosmic time with respect to us of 1 billion year. Now 
one may ask what will be the Big Bang time with respect to this galaxy. Will
it be the Big Bang time with respect to us minus 1 billion year? A person
who lives in that galaxy will look at our galaxy and say that ours is far away 
from him by also 1 billion year. Will that mean, with respect to him, our
galaxy is closer to the Big Bang time by 1 billion year? Or will we seem to 
him to be farther by 1 billion year? All this leads to the conclusion that 
there is no absolute cosmic time. Rather, it is a relative concept.

Based on this assumption, and using the Hubble expansion $R=\tau v$, where $R$
is the distance from us to a galaxy, $v$ is the receding velocity of the 
galaxy and $\tau$ is the Hubble time (a universal constant equal
to 12.486 Gyr - it is the standard Hubble time in the limit of zero distance
and zero gravity - and thus it is a constant in this epoch of time, see 
Section 8). The 
Hubble expansion can now be written as 
\begin{equation}
\tau^2 v^2-\left(x^2+y^2+z^2\right)=0.
\label{eq.1}
\end{equation}
($\tau v$ can alternatively be written in terms of the redshift. For 
nonrelativistic and relativistic velocities the relationship between the 
velocity and the redshift parameter are given respectively by $z=v/c$ and 
$z=[(1+v/c)/(1-v/c)]^{1/2}-1$, $v/c=z(z+2)/(z^2+2z+2)$.)  
Accordingly we have a line element 
\begin{equation}
ds^2=\tau^2 dv^2-\left(dx^2+dy^2+dz^2\right).
\label{eq.2} 
\end{equation}
It is equal to zero for the Hubble expansion, but is otherwise not vanishing 
for cosmic times smaller than $\tau$. 
It is similar to the Minkowskian metric
\begin{equation}
ds^2=c^2dt^2-\left(dx^2+dy^2+dz^2\right),
\label{eq.3}
\end{equation}
which vanishes for light propagation, but is otherwise different from zero for
particles of finite mass.

If we now assume that the laws of physics are valid at all cosmic times and
$\tau$ is a universal constant which has the same value at all cosmic times,
then we can develop a new special relativity just like Einstein's original
special relativity theory: the validity of the laws of physics at all cosmic 
times replaces the special relativistic assumption in Einstein's theory of in
all inertial coordinate systems, whereas the constancy of $\tau$ at all cosmic 
times replaces the constancy of the speed of light in all inertial systems in
ordinary special relativity. 

In this way one also obtains a cosmological transformation that relates 
distances and velocities (redshifts) at different cosmic times, just as in
ordinary special relativity we have the Lorentz transformation that relates
spatial coordinates and time at different velocities. We now
will have
\begin{equation}
x'=\frac{x-tv}{\sqrt{1-t^2/\tau^2}},
\hspace{5mm}
v'=\frac{v-xt/\tau^2}{\sqrt{1-t^2/\tau^2}},
\label{eq.4}
\end{equation}
for the case $y'=y$, $z'=z$. In the above transformation $t$ is the relative
cosmic time and is measured backward ($t=0$ now and $t=\tau$ at the Big Bang). 
As seen from the above transformation, $\tau$ is the maximum 
cosmic time that is allowed in nature and as such can be considered as the age 
of the Universe.

For example, if we denote the temperature at a cosmic time $t$ by $T$, and the
temperature at the present time by $T_0$ (=2.73K), we then have 
$T=T_0/\sqrt{1-t^2/\tau^2}$. For instance, at $t/\tau=1/2$ we get 
$T=2\times 2.73/\sqrt{3}=
3.15$K. (This result does not take into account gravity that needs a
correction by a factor of
13, and thus the temperature at $t=\tau/2$ is 40.95K.) 

Since we always use forward rather than backward times, we write the 
transformation~(\protect\ref{eq.4}) in terms of such a time $t'=\tau-t$. The resulting 
transformation will have the form 
\begin{equation}
x'=\frac{x-\left(\tau-t\right)v}{\sqrt{\left(t/\tau\right)\left(2-t/\tau\right)}},
\hspace{5mm}
v'=\frac{v-x\left(\tau-t\right)/\tau^2}{\sqrt{\left(t/\tau\right)\left(2-t/\tau\right)}},
\label{eq.5}
\end{equation}
where primes have been dropped for brevity, $0\leq t\leq\tau$, $t=0$ at the Big 
Bang, $t=\tau$, now.

The above introduction gives a brief review of a new special relativity 
(cosmological special relativity, for more details see~\cite{1}). Obviously 
the Universe is filled up with
gravity and therefore one has to go to a Riemannian space with the Einstein
gravitational field equations in terms of space and redshift (velocity). This
is done in  the next section.
\section{Extension to curved space: cosmological general relativity}
The theory presented here, cosmological general relativity, uses a Riemannian 
four-dimensional presentation of 
gravitation in which the coordinates are those of Hubble, i.e. distances and 
velocity rather than the traditional space and time. We solve the field 
equations and show
that there are three possibilities for the Universe to expand. The theory 
describes the Universe as having a three-phase evolution with a decelerating
expansion, followed by a constant and  an accelerating expansion, and it 
predicts that the Universe is now in the latter phase. It is
shown, assuming $\Omega_m=0.245$, that the time at which the Universe goes over
from a decelerating to an accelerating expansion, i.e., the constant-expansion
phase, occurs at 8.5 Gyr ago. Also, at that time the cosmic radiation 
temperature was 146K. Recent
observations of distant supernovae imply, in defiance of expectations, that 
the Universe's growth is accelerating, contrary to what has always been 
assumed, that the expansion is slowing down due to gravity. Our theory 
confirms these recent experimental results by showing that the Universe now is 
definitely in a stage of accelerating expansion. The theory predicts also that 
now there is a positive pressure, $p=0.034g/cm^2$, in the Universe. Although 
the theory has no
cosmological constant, we extract from it its equivalence and show that 
$\Lambda=1.934\times 10^{-35}s^{-2}$. This value of 
$\Lambda$ is in excellent agreement with the measurements obtained by the
{\it High-Z Supernova Team} and the {\it Supernova Cosmology Project}. It is 
also shown that the three-dimensional space of the Universe is Euclidean, as 
the Boomerang, Maxima, DASI and CBI microwave telescopes have shown. 
Comparison with general relativity theory is
finally made and it is pointed out that the classical experiments as well as the 
gravitational radiation prediction follow from the present theory, too.
\section{Cosmology in spacevelocity}
In the 
framework of cosmological general relativity (CGR) gravitation is described
by a curved four-dimensional Riemannian spacevelocity~\cite{2,3,4,5}. CGR incorporates the
Hubble constant $\tau$ at the outset. The Hubble law is assumed in CGR as a
fundamental law. CGR, in essence, extends Hubble's law so as to incorporate 
gravitation in it; it is actually a {\it distribution theory} that relates distances 
and velocities between galaxies. The theory involves only measured quantities
and it takes a picture of the Universe as it is at any moment. The following 
is a brief review of CGR as was originally given by the author in 1996 in Ref. 
2.

The foundations of any gravitational theory are based on the principle of
equivalence and the principle of general covariance~\cite{6}. These two principles lead immediately
to the realization that gravitation should be described by a four-dimensional
curved spacetime, in our theory spacevelocity, and that the field equations 
and the equations of motion should be written in a generally covariant form.
Hence these principles were adopted in CGR also. Use is made in a 
four-dimensional Riemannian manifold with a metric $g_{\mu\nu}$ and a line 
element $ds^2=g_{\mu\nu}dx^\mu dx^\nu$. The difference from Einstein's general
relativity is that our coordinates are: $x^0$ is a velocitylike coordinate 
(rather than a timelike coordinate), thus $x^0=\tau v$ where $\tau$ is the
Hubble time in the zero-gravity limit and $v$ the velocity. The coordinate 
$x^0=\tau v$ is the 
comparable to $x^0=ct$ where $c$ is the speed of light and $t$ is the time in
ordinary general relativity. The other three coordinates $x^k$, $k=1,2,3$, are
spacelike, just as in general relativity theory.

An immediate consequence of the above choice of coordinates is that the null
condition $ds=0$ describes the expansion of the Universe in the curved 
spacevelocity (generalized Hubble's law with gravitation) as compared to the 
propagation of light in the curved spacetime in general relativity. This means
one solves the field equations (to be given in the sequel) for the metric 
tensor, then from the null condition $ds=0$ one obtains immedialety the 
dependence of the relative distances between the galaxies on their relative
velocities.

As usual in gravitational theories, one equates geometry to physics. The first 
is expressed by means of a combination of the Ricci tensor and the Ricci
scalar, and follows to be naturally either the Ricci trace-free tensor or the
Einstein tensor. The Ricci trace-free tensor does not fit gravitation in 
general, and
the Einstein tensor is a natural candidate. The physical part is expressed by
the energy-momentum tensor which now has a different physical meaning from 
that in Einstein's theory. More important, the coupling constant that relates 
geometry to physics is now also {\it different}. 

Accordingly the field equations are
\begin{equation}
G_{\mu\nu}=R_{\mu\nu}-\frac{1}{2}g_{\mu\nu}R=\kappa T_{\mu\nu},
\label{eq.6}
\end{equation}
exactly as in Einstein's theory, with $\kappa$ given by
$\kappa=8\pi k/\tau^4$, (in general relativity it is given by $8\pi G/c^4$),
where $k$ is given by $k=G\tau^2/c^2$, with $G$ being Newton's gravitational
constant, and $\tau$ the Hubble constant time. When the equations of motion 
will be written in terms of velocity instead of time, the constant $k$ will
replace $G$. Using the above equations one then has $\kappa=8\pi G/c^2\tau^2$.

The energy-momentum tensor $T^{\mu\nu}$ is constructed, along the lines of
general relativity  theory, with the speed of light being replaced by the
Hubble constant time. If $\rho$ is the average mass density of the Universe,
then it will be assumed that $T^{\mu\nu}=\rho u^\mu u^\nu,$ where $u^\mu=
dx^\mu/ds$ is the four-velocity.
In general relativity theory one takes $T_0^0=\rho$. In Newtonian gravity one
has the Poisson equation $\nabla^2\phi=4\pi G\rho$. At points where $\rho=0$
one solves the vacuum Einstein field equations in general relativity and the 
Laplace equation 
$\nabla^2\phi=0$ in Newtonian gravity. In both theories a null (zero) solution
is allowed as a trivial case. In cosmology, however, there exists no situation
at which $\rho$ can be zero because the Universe is filled with matter. In
order to be able to have zero on the right-hand side of~(\protect\ref{eq.3}) 
one takes 
$T_0^0$ not as equal to $\rho$, but to $\rho_{eff}=\rho-\rho_c$, where 
$\rho_c$ is the critical mass density, 
a {\it constant} in CGR given by $\rho_c=3/8\pi G\tau^2$, whose value is 
$\rho_c\approx 10^{-29}g/cm^3$, a few hydrogen atoms per cubic meter. 
Accordingly one takes
$T^{\mu\nu}=\rho_{eff}u^\mu u^\nu$; $\rho_{eff}=\rho-\rho_c$
for the energy-momentum tensor. Moreover, the above choice of the 
energy-momentum tensor is the only possibility that yields a constant
expansion when $\rho=\rho_c$ as it should be (see Section 6).

In the next sections we apply CGR to obtain the accelerating expanding 
Universe and related subjects.
\section{Gravitational field equations}
In the four-dimensional spacevelocity the spherically symmetric metric is 
given by
\begin{equation}
ds^2=\tau^2dv^2-e^\mu dr^2-R^2\left(d\theta^2+\sin^2\theta d\phi^2\right),
\label{eq.7}
\end{equation}
where $\mu$ and $R$ are functions of $v$ and $r$ alone, and comoving 
coordinates $x^\mu=(x^0,x^1,x^2,x^3)=(\tau v,r,\theta,\phi)$ have been used. 
With the above choice of coordinates, the zero-component of the geodesic
equation becomes an identity, and since $r$, $\theta$ and $\phi$ are constants
along the geodesics, one has $dx^0=ds$ and therefore
$u^\alpha=u_\alpha=\left(1,0,0,0\right).$
The metric~(\protect\ref{eq.7}) shows that the area of the sphere $r=constant$ 
is given by
$4\pi R^2$ and that $R$ should satisfy $R'=\partial R/\partial r>0$. The
possibility that $R'=0$ at a point $r_0$ is excluded since it would
allow the lines $r=constants$ at the neighboring points $r_0$ and $r_0+dr$ to
coincide at $r_0$, thus creating a caustic surface at which the comoving 
coordinates break down.

As has been shown in the previous sections the Universe expands by the null
condition $ds=0$, and if the expansion is spherically symmetric one has
$d\theta=d\phi=0$. The metric~(\protect\ref{eq.7}) then yields
$\tau^2 dv^2-e^\mu dr^2=0,$
thus
\begin{equation}
\frac{dr}{dv}=\tau e^{-\mu/2}.
\label{eq.8}
\end{equation}
This is the differential equation that determines the Universe expansion. In
the following we solve the gravitational field equations in order to find out
the function $\mu\left(r.v\right)$.

The gravitational field equations ~(\protect\ref{eq.6}), written in the form
\begin{equation}
R_{\mu\nu}=\kappa\left(T_{\mu\nu}-g_{\mu\nu}T/2\right),
\label{eq.9}
\end{equation}
where 
\begin{equation} 
T_{\mu\nu}=\rho_{eff}u_\mu u_\nu+p\left(u_\mu u_\nu-g_{\mu\nu}\right), 
\label{eq.10}
\end{equation} 
with $\rho_{eff}=\rho-\rho_c$ and $T=T_{\mu\nu}g^{\mu\nu}$, are now solved.
One finds that the only nonvanishing components of $T_{\mu\nu}$ 
are $T_{00}=\tau^2\rho_{eff}$, $T_{11}=c^{-1}\tau pe^\mu$, $T_{22}=c^{-1}\tau 
pR^2$ and $T_{33}=c^{-1}\tau pR^2\sin^2\theta$, and that $T=\tau^2\rho_{eff}-
3c^{-1}\tau p$.

The field equations obtained are given by
\begin{equation}
-\ddot{\mu}-\frac{4}{R}\ddot{R}-\frac{1}{2}\dot{\mu}^2=\kappa\left(
\tau^2\rho_{eff}+3c^{-1}\tau p\right),
\label{eq.11}
\end{equation}
\begin{equation}
2\dot{R}'-R'\dot{\mu}=0,
\label{eq.12} 
\end{equation}
\begin{equation}
\ddot{\mu}+\frac{1}{2}\dot{\mu}^2+\frac{2}{R}\dot{R}\dot{\mu}+e^{-\mu}\left(
\frac{2}{R}R'\mu'-\frac{4}{R}R''\right)=\kappa\left(\tau^2\rho_{eff}-c^{-1}\tau 
p\right)
\label{eq.13}
\end{equation}
\begin{equation}
\frac{2}{R}\ddot{R}+2\left(\frac{\dot{R}}{R}\right)^2+\frac{1}{R}\dot{R}
\dot{\mu}+\frac{2}{R^2}+e^{-\mu}\left[\frac{1}{R}R'\mu'-2\left(\frac{R'}{R}
\right)^2-\frac{2}{R}R''\right]\\
=\kappa\left(\tau^2\rho_{eff}-c^{-1}\tau p\right).
\label{eq.14}
\end{equation}
Combinations of 
Eqs. ~(\protect\ref{eq.11})--~(\protect\ref{eq.14}) then give three independent field equations:
\begin{equation}
e^\mu\left(2R\ddot{R}+\dot{R}^2+1\right)-R'^2=-\kappa\tau c^{-1} e^\mu R^2p,
\label{eq.15}
\end{equation}
\begin{equation}
2\dot{R}'-R'\dot{\mu}=0, 
\label{eq.16}
\end{equation}
\begin{equation}
e^{-\mu}\left[\frac{1}{R}R'\mu'-\left(\frac{R'}{R}\right)^2-\frac{2}{R}R''
\right]+\frac{1}{R}\dot{R}\dot{\mu}+\left(\frac{\dot{R}}{R}\right)^2+
\frac{1}{R^2}=\kappa\tau^2\rho_{eff}, 
\label{eq.17}
\end{equation}
other equations being trivial combinations 
of~(\protect\ref{eq.15})--~(\protect\ref{eq.17}).
\section{Solution of the field equations}
The solution of~(\protect\ref{eq.16}) satisfying the condition $R'>0$ is given by
\begin{equation}
e^\mu=R'^2/\left(1+f\left(r\right)\right),
\label{eq.18}
\end{equation}
where $f\left(r\right)$ is an arbitrary function of the coordinate $r$ and 
satisfies the
condition $f\left(r\right)+1>0$. Substituting~(\protect\ref{eq.18}) in the other two 
field equations~(\protect\ref{eq.15}) and~(\protect\ref{eq.17}) then gives
\begin{equation}
2R\ddot{R}+\dot{R}^2-f=-\kappa c^{-1}\tau R^2p, 
\label{eq.19}
\end{equation}
\begin{equation}
\frac{1}{RR'}\left(2\dot{R}\dot{R'}-f'\right)+\frac{1}{R^2}\left(\dot{R}^2-f
\right)=\kappa\tau^2\rho_{eff},
\label{eq.20}
\end{equation}
respectively.

The simplest solution of the above two equations, which satisfies the 
condition $R'=1>0$, is given by $R=r$.
Using this in Eqs.~(\protect\ref{eq.19}) and~(\protect\ref{eq.20}) gives
$f\left(r\right)=\kappa c^{-1}\tau pr^2$, and
$f'+f/r=-\kappa\tau^2\rho_{eff}r$,
respectively. 
Using the values of $\kappa=8\pi G/c^2\tau^2$ and $\rho_c=3/8\pi G\tau^2$, we
obtain
\begin{equation}
f\left(r\right)=\left(1-\Omega_m\right)r^2/\left(c^2\tau^2\right),
\label{eq.21}
\end{equation}
where $\Omega_m=\rho/\rho_c$. We also obtain
\begin{equation} 
p=\frac{1-\Omega_m}{\kappa c\tau^3}=\frac{c}{\tau}\frac{1-\Omega_m}{8\pi G}
=4.544\left(1-\Omega_m\right)\times 10^{-2} g/cm^2,
\label{eq.22}
\end{equation}
\begin{equation}
e^{-\mu}=1+f\left(r\right)=1+\tau c^{-1}\kappa pr^2=1+
\left(1-\Omega_m\right)r^2/\left(c^2\tau^2 \right).
\label{eq.23}
\end{equation} 

Accordingly, the line element of the Universe is given by
\begin{equation}
ds^2=\tau^2dv^2-\frac{dr^2}{1+\left(1-\Omega\right)r^2/\left(c^2\tau^2\right)}
-r^2\left(d\theta^2+\sin^2\theta d\phi^2\right),
\end{equation}
or,
\begin{equation}
ds^2=\tau^2dv^2-\frac{dr^2}{1+\left(\kappa\tau/c\right)pr^2}
-r^2\left(d\theta^2+\sin^2\theta d\phi^2\right).
\label{eq.23a}
\end{equation}
This line element is the comparable to the FRW line element in the standard theory.

It  will be recalled that the Universe expansion is determined 
by Eq.~(\protect\ref{eq.8}),
$dr/dv=\tau e^{-\mu/2}$. The only thing that is left to be determined is the
signs of $(1-\Omega_m)$ or the pressure $p$. Thus we have
\begin{equation}
\frac{dr}{dv}=\tau\sqrt{1+\kappa\tau c^{-1}pr^2}=\tau\sqrt{1+
\frac{1-\Omega_m}{c^2\tau^2}r^2}.
\label{eq.24}
\end{equation} 
For simplicity we confine ourselves to the linear approximation, 
thus Eq.~(\protect\ref{eq.24}) yields
\begin{equation}
\frac{dr}{dv}=\tau\left(1+\frac{\kappa}{2}\tau c^{-1}pr^2\right)=
\tau\left[1+\frac{1-\Omega_m}{2c^2\tau^2}r^2\right].
\label{eq.25}
\end{equation}
\section{Physical meaning}
To see the physical meaning of these solutions, one does not need the
exact solutions. Rather, it is enough to write down the solutions in the
lowest approximation in $\tau^{-1}$. One obtains, by 
differentiating Eq.~(\protect\ref{eq.25}) with respect to $v$, for $\Omega_m>1$,
\begin{equation}
d^2r/dv^2=-kr;\mbox{\hspace{10mm}}k=\left(\Omega_m-1\right)/\left(2c^2\right),
\label{eq.26}
\end{equation}
the solution of which is
\begin{equation} 
r\left(v\right)=A\sin\alpha\frac{v}{c}+B\cos\alpha\frac{v}{c},
\label{eq.27}
\end{equation}
where $\alpha^2=(\Omega_m-1)/2$ and $A$ and $B$ are constants. The latter can be
determined by the initial condition $r\left(0\right)=0=B$ and $dr\left(0
\right)/dv=\tau=A\alpha/c$, thus
\begin{equation}
r\left(v\right)=\frac{c\tau}{\alpha}\sin\alpha\frac{v}{c}.
\label{eq.28}
\end{equation}
This is obviously a closed Universe, and presents a decelerating expansion.

For $\Omega_m<1$ we have
\begin{equation}
d^2r/dv^2=\left(1-\Omega_m\right)r/\left(2c^2\right),
\label{eq.29}
\end{equation}
whose solution, using the same initial conditions, is
\begin{equation}
r\left(v\right)=\frac{c\tau}{\beta}\sinh\beta\frac{v}{c},
\label{eq.30}
\end{equation}
where $\beta^2=(1-\Omega_m)/2$. This is now an open accelerating Universe.

For $\Omega_m=1$ we have, of course, $r=\tau v$.
\section{The accelerating universe}
We finally determine which of the three cases of expansion is the one at 
present epoch of time. To this end we have to write the 
solutions~(\protect\ref{eq.28}) and~(\protect\ref{eq.30}) in ordinary Hubble's 
law form $v=H_0r$. Expanding Eqs.~(\protect\ref{eq.28}) 
and~(\protect\ref{eq.30}) into power series in $v/c$ and keeping terms up to 
the second order, we obtain
\begin{equation}
r=\tau v\left(1-\alpha^2v^2/6c^2\right)=\tau v\left(1+\beta^2v^2/6c^2\right), 
\label{eq.31}
\end{equation}
for $\Omega_m>1$ and $\Omega_m<1$. Using now the expressions for 
$\alpha$ and $\beta$, then Eqs.~(\protect\ref{eq.31}) reduce to the single 
equation
\begin{equation}
r=\tau v\left[1+\left(1-\Omega_m\right)v^2/6c^2\right].
\label{eq.32}
\end{equation}
Inverting now this equation by writing it as $v=H_0r$, we obtain in the lowest
approximation
\begin{equation}
H_0=h\left[1-\left(1-\Omega_m\right)v^2/6c^2\right],
\label{eq.33}
\end{equation}
where $h=\tau^{-1}$. To the same approximation one also obtains
\begin{equation}
H_0=h\left[1-\left(1-\Omega_m\right)z^2/6\right]=h\left[1-\left(1-\Omega_m
\right)r^2/6c^2\tau^2\right],
\label{eq.34}
\end{equation}
where $z$ is the redshift parameter.
As is seen, and it is confirmed by experiments, $H_0$ depends on the distance 
it is being measured; it has physical meaning only at the zero-distance limit,
namely when measured {\it locally}, in which case it becomes $h=1/\tau$.

It is well known that the measured value of $H_0$ depends on the ``short"
and ``long" distance scales~\cite{8}. The farther the distance $H_0$ is being 
measured, the lower the value for $H_0$ is obtained. By Eq.~(\protect\ref{eq.34}) 
this is
possible only when $\Omega_m<1$, namely when the Universe is accelerating. By
Eq.~(\protect\ref{eq.22}) we also find that the pressure is positive. 

The possibility that the Universe expansion is accelerating was first 
predicted using CGR by the author in 1996~\cite{2} before the supernovae 
experiments results became known.

It  will be noted that the constant expansion is just a transition stage 
between the decelerating and the accelerating expansions as the Universe
evolves toward its present situation.

\begin{table}[h]
\begin{center} 
\begin{tabular}{c r@{}l c r@{.}l r@{}l r@{}l}
\hline\\ 
Curve&\multicolumn{2}{c}{$\Omega_m$}&Time in Units &\multicolumn{2}{c}{Time}&
\multicolumn{2}{c}{Temperature}&\multicolumn{2}{c}{Pressure}\\
No$^\star$.& & & of $\tau$ &\multicolumn{2}{c}{(Gyr)}& &(K)&
\multicolumn{2}{c}{(g/cm$^2$)}\\
\hline\\
\multicolumn{10}{c}{DECELERATING EXPANSION}\\
1&100& &$3.1\times 10^{-6}$&$3$&$87\times 10^{-5}$&1096&&-&4.499\\
2&25& &$9.8\times 10^{-5}$&$1$&$22\times 10^{-3}$&195&.0&-&1.091\\
3&10& &$3.0\times 10^{-4}$&$3$&$75\times 10^{-3}$&111&.5&-&0.409\\
4&5& &$1.2\times 10^{-3}$&$1$&$50\times 10^{-2}$&58&.20&-&0.182\\
5&1&.5&$1.3\times 10^{-2}$&$1$&$62\times 10^{-1}$&16&.43&-&0.023\\
\multicolumn{10}{c}{CONSTANT EXPANSION}\\ 
6&1& &$3.0\times 10^{-2}$&$3$&$75\times 10^{-1}$&11&.15&&0\\
\multicolumn{10}{c}{ACCELERATING EXPANSION}\\ 
7&0&.5&$1.3\times 10^{-1}$&$1$&$62$&5&.538&+&0.023\\
8&0&.245&$1.0$&$12$&$50$&2&.730&+&0.034\\
\hline\\ 
\end{tabular}
\caption{The Cosmic Times with respect to the Big Bang, the Cosmic
Temperature and the Cosmic Pressure for each of the Curves in Fig. 1.
\label{t1}}
\end{center}
\end{table}

Figure 1 describes the Hubble diagram of the above solutions for the three 
types of expansion for values of $\Omega_m$ from 100 to 0.245. The figure
describes the three-phase evolution of the Universe. Curves (1)-(5) represent
the stages of {\it decelerating expansion} according to 
Eq.~(\protect\ref{eq.28}). As the density 
of matter $\rho$ decreases, the Universe goes over from the lower curves to 
the upper ones, but it does not have enough time to close up to a big crunch.
The Universe subsequently goes over to curve (6) with $\Omega_m=1$, at which time
it has a constant expansion for a fraction of a second. This then followed
by going to the upper curves (7) and (8) with $\Omega_m<1$, where the Universe
expands with {\it acceleration} according to Eq.~(\protect\ref{eq.30}). Curve 
no. 8 fits
the present situation of the Universe. For curves (1)-(4) in the diagram we
use the cutoff when the curves were at their maximum. In Table 1 we present 
the cosmic times with respect to the big bang, the cosmic radiation
temperature and the pressure for each of the curves in Fig. 1.

\begin{figure}[t]
\centering
\includegraphics[width=10cm]{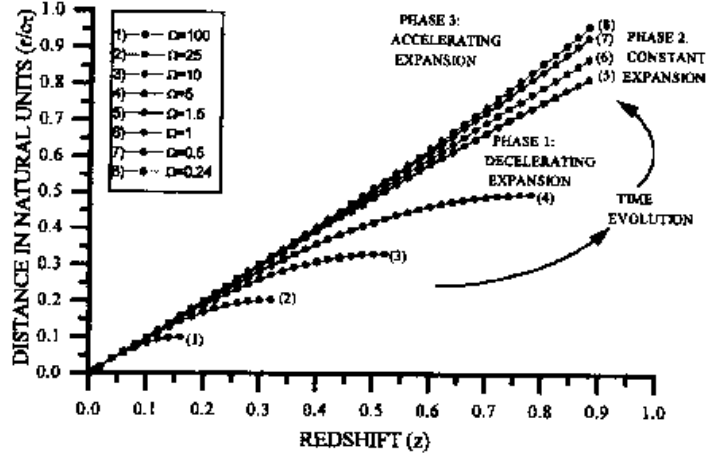}
\caption{Hubble's diagram describing the three-phase evolution of the
Universe according to cosmological general relativity theory. Curves (1) to
(5) represent the stages of {\it decelerating} expansion. As the density of 
matter $\rho$ decreases, the Universe goes over
from the lower curves to the upper ones, but it does not have enough time to
close up to a big crunch. The Universe subsequently goes to curve (6) with
$\Omega=1$, at which time it has a {\it constant} expansion for a fraction of
a second. This then followed by going to the upper curves (7), (8) with 
$\Omega<1$ where the Universe expands with {\it acceleration}. Curve no. 8
fits the present situation of the Universe. (Source: S. Behar and M. Carmeli,
Ref. 3)}
\label{fig.1}
\end{figure}

Figure 2 shows the Hubble diagrams for the distance-redshift 
relationship predicted by the theory for the accelerating expanding Universe 
at the present time, and Figure 3 shows the experimental results.
\begin{figure}[t]
\centering
\includegraphics[width=10cm]{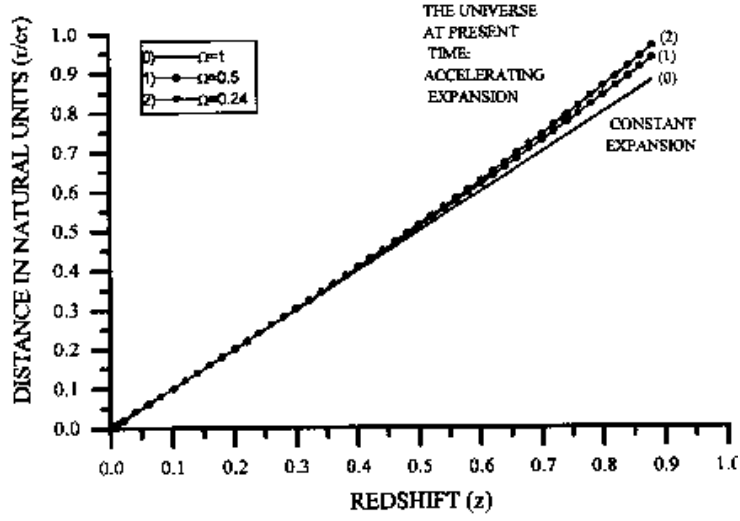}
\caption{Hubble's diagram of the Universe at the present phase of evolution
with accelerating expansion. (Source: S. Behar and M. Carmeli,
Ref. 3)}
\end{figure}
\begin{figure}[t]
\centering
\includegraphics[width=10cm]{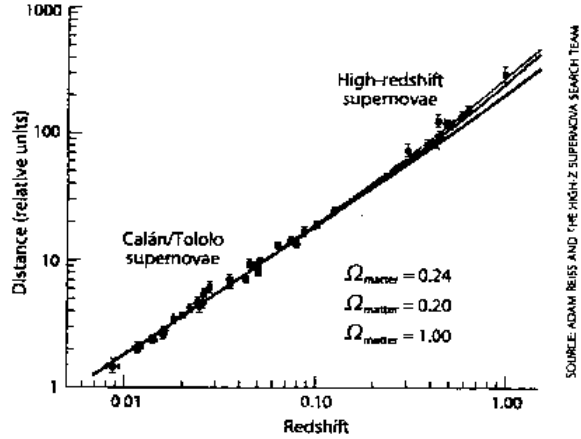}
\caption{Distance vs. redshift diagram showing the deviation from a constant
toward an accelerating expansion. (Source: A. Riess {\it et al.}, Ref. 12)}
\end{figure}

Our estimate for $h$, based on published data, is $h\approx 80$ km/sec-Mpc.
Assuming $\tau^{-1}\approx 80$ km/sec-Mpc, Eq.~(\protect\ref{eq.34}) then 
gives
\begin{equation}
H_0=h\left[1-1.3\times 10^{-4}\left(1-\Omega_m\right)r^2\right],
\label{eq.35}
\end{equation}
where $r$ is in Mpc. A computer best-fit can then fix both $h$ and $\Omega_m$.

To summarize, a theory of cosmology has been presented in which the dynamical
variables are those of Hubble, i.e. distances and velocities. The theory
descirbes the Universe as having a three-phase evolution with a decelerating 
expansion, followed by a constant and an accelerating expansion, and it
predicts that the Universe is now in the latter phase. As the density of 
matter decreases, while the Universe is at the decelerating phase, it does not
have enough time to close up to a big crunch. Rather, it goes to the 
constant-expansion phase, and then to the accelerating stage. As we have seen,
the equation obtained for the Universe expansion, Eq.~(\protect\ref{eq.30}), 
is very simple.
\section{Theory versus experiment}
The Einstein gravitational 
field equations  
with the added cosmological term are~\cite{9}:
\begin{equation}
R_{\mu\nu}-\frac{1}{2}g_{\mu\nu}R+\Lambda g_{\mu\nu}=\kappa T_{\mu\nu},
\label{eq.36}
\end{equation}
where $\Lambda$ is the cosmological constant, the value of which is supposed to
be determined by the theory of quantum gravity which nobody knows much about it. 
In Eq.~(\protect\ref{eq.36}) $R_{\mu\nu}$ and $R$ are the Ricci 
tensor and scalar, respectively, $\kappa=8\pi G$, where $G$ is Newton's constant
and the speed of light is taken as unity.

Recently the two groups (the {\it Supernovae Cosmology Project} and the {\it 
High-Z Supernova Team}) concluded that the expansion of the Universe is 
accelerating~\cite{10,11,12,13,14,15,16}. The two groups had discovered and measured moderately 
high redshift ($0.3<z<0.9$) supernovae, and found that they were fainter than
what one would expect them to be if the cosmos expansion were slowing down or
constant. Both teams obtained $\Omega_m\approx 0.3$, $\Omega_\Lambda\approx 
0.7$, and ruled out the traditional ($\Omega_m$, $\Omega_\Lambda$)=(1, 0)
Universe. Their value of the density parameter $\Omega_\Lambda$ corresponds to
a cosmological constant that is small but, nevertheless, nonzero and positive,
$\Lambda\approx 10^{-52}\mbox{\rm m}^{-2}\approx 10^{-35}\mbox{\rm s}^{-2}$.

In previous sections a four-dimensional cosmological theory (CGR) was 
presented. Although the theory has no cosmological constant, it predicts that 
the Universe accelerates and hence it has the equivalence of a positive 
cosmological constant in 
Einstein's general relativity. In the framework of this theory (see Section 
3) the 
zero-zero component of the field equations~(\protect\ref{eq.6}) is written as
\begin{equation}
R_0^0-\frac{1}{2}\delta_0^0R=\kappa\rho_{eff}=\kappa\left(\rho-\rho_c
\right),
\label{eq.37}
\end{equation}
where $\rho_c=3/\kappa\tau^2$ is the critical mass density
and $\tau$ is Hubble's time in the zero-gravity limit.

Comparing Eq.~(\protect\ref{eq.37}) with the zero-zero component of 
Eq.~(\protect\ref{eq.36}), one obtains 
the expression for the cosmological constant of general relativity, 
$\Lambda=\kappa\rho_c=3/\tau^2.$

To find out the numerical value of $\tau$ we use the relationship between
$h=\tau^{-1}$ and $H_0$ given by Eq.~(\protect\ref{eq.34}) (CR denote values 
according to
Cosmological Relativity):
\begin{equation} 
H_0=h\left[1-\left(1-\Omega_m^{CR}\right)z^2/6\right],
\label{eq.38}
\end{equation}
where $z=v/c$ is the redshift and $\Omega_m^{CR}=\rho_m/\rho_c$ with 
$\rho_c=3h^2/8\pi G$. (Notice that our $\rho_c=1.194\times 10^{-29}g/cm^3$ is different from 
the standard $\rho_c$ defined with $H_0$.) The redshift parameter $z$ 
determines the distance at which $H_0$ is measured. We choose $z=1$ and take 
for $\Omega_m^{CR}=0.245$,
its value at the present time (see Table 1) (corresponds to 0.32 in the 
standard theory), Eq.~(\protect\ref{eq.38}) then gives $H_0=0.874h.$
At the value $z=1$ the corresponding Hubble parameter $H_0$ according to the 
latest results from HST can be taken~\cite{17} as $H_0=70$km/s-Mpc, thus 
$h=(70/0.874)$km/s-Mpc, or $h=80.092\mbox{\rm km/s-Mpc},$
and $\tau=12.486 Gyr=3.938\times 10^{17}s.$

What is left is to find the value of $\Omega_\Lambda^{CR}$. We have 
$\Omega_\Lambda^{CR}=\rho_c^{ST}/\rho_c$, where $\rho_c^{ST}=3H_0^2/8\pi 
G$ and $\rho_c=3h^2/8\pi G$. Thus $\Omega_\Lambda^{CR}=(H_0/h)^2=0.874^2$,
or $\Omega_\Lambda^{CR}=0.764.$
As is seen from the above equations one has 
\begin{equation}
\Omega_T=\Omega_m^{CR}+\Omega_\Lambda^{CR}=0.245+0.764=1.009\approx 1,
\label{eq.39}
\end{equation}
which means the Universe is Euclidean.

As a final result we calculate the cosmological constant. One obtains
\begin{equation}
\Lambda=3/\tau^2=1.934\times 10^{-35}s^{-2}.
\label{eq.40}
\end{equation}

Our results confirm those of the supernovae experiments and indicate on the
existance of the dark energy as has recently received confirmation from the
Boomerang cosmic microwave background experiment~\cite{18,19}, which showed that 
the Universe is Euclidean.
\section{Remarks}
In this paper the cosmological general relativity, a relativistic theory in
spacevelocity, has been presented and applied to the problem of the expansion 
of the Universe. The theory, which predicts a positive pressure for the 
Universe now, describes the Universe as having a three-phase
evolution: decelerating, constant and accelerating expansion, but it is now in
the latter stage. Furthermore, the cosmological constant that was
extracted from the theory agrees with the experimental result. Finally, it has
also been shown that the three-dimensional spatial space of the Universe is
Euclidean, again in agreement with observations. 

Recently~\cite{20,21}, more confirmation to the Universe accelerating expansion 
came from 
the most distant supernova, SN 1997ff, that was recorded by the Hubble Space
Telescope. As has been pointed out before, if we look back far enough, we 
should find a decelerating expansion (curves 1-5 in Figure 1). Beyond $z=1$
one should see an earlier time when the mass density was dominant. The 
measurements obtained from SN 1997ff's redshift and brightness provide a 
direct proof for the transition from past decelerating to present 
accelerating expansion (see Figure 4). The measurements also exclude 
the possibility that
the acceleration of the Universe is not real but is due to other astrophysical 
effects such as dust.
\begin{figure}[t]
\centering
\includegraphics[height=10cm,angle=90]{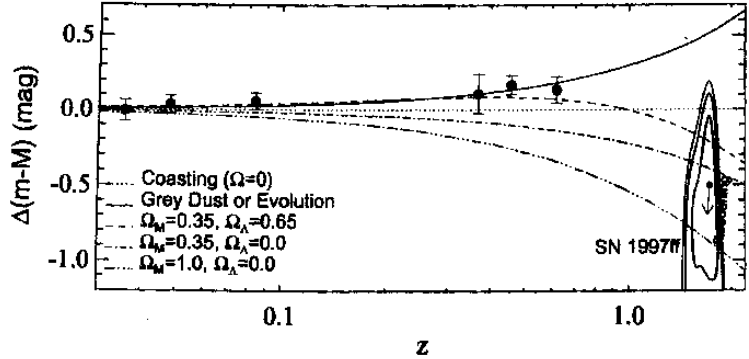}
\caption{Hubble diagram of SNe Ia minus an empty (i.e., ``empty" $\Omega=0$)
Universe compared to cosmological and astrophysical models. The points are
the redshift-binned data from the HZT (Riess {\it et al.} 1998) and the SCP
(Perlmutter {\it et al.} 1999). The measurements of SN 1997ff are inconsistent
with astrophysical effects which could mimic previous evidence for an 
accelerating Universe from SNe Ia at $z\approx 0.5$. (Source: A. Riess 
{\it et al.}, Ref. 21)}
\label{fig.4}
\end{figure}

Table 2 gives some of the cosmological parameters obtained here and in the
standard theory.

\begin{table}[h]
\begin{center}  
\begin{tabular}{p{35mm}p{35mm}p{35mm}}
\hline\\
&COSMOLOGICAL&STANDARD\\
&RELATIVITY&THEORY\\
\hline\\
Theory type&Spacevelocity&Spacetime\\
Expansion&Tri-phase:&One phase\\
type&decelerating, constant,&\\
&accelerating&\\
Present expansion&Accelerating&One of three\\
&(predicted)&possibilities\\
Pressure&$0.034g/cm^2$&Negative\\
Cosmological constant&$1.934\times 10^{-35}s^{-2}$&Depends\\
&(predicted)&\\
$\Omega_T=\Omega_m+\Omega_\Lambda$&1.009&Depends\\
Constant-expansion&8.5Gyr ago&No prediction\\
occurs at&(Gravity is included)&\\
Constant-expansion&Fraction of&Not known\\
duration&second&\\
Temperature at&146K&No prediction\\
constant expansion&(Gravity is included)&\\
\hline
\end{tabular}
\caption{Cosmological parameters in cosmological general  relativity and in
standard theory.   \label{t2}}
\end{center}
\end{table}

In order to compare the present theory with general relativity, we must add the 
time coordinate. We then have a time-space-velocity Universe with two 
time-like and three space-like coordinates, with signature $(+---+)$. We are
concerned with the classical experiments of general relativity and the 
gravitational waves predicted by that theory. It can be shown that
all these results are also obtained from the present theory~\cite{22}.  

In the case of gravitational waves we get in this theory a more general
formula than that obtained in general relativity theory. Writing the metric 
$g_{\mu\nu}\approx\eta_{\mu\nu}+h_{\mu\nu}$, where $h_{\mu\nu}$ is a 
first-approximation
term, and using the notation $h_{\mu\nu}=\gamma_{\mu\nu}-\eta_{\mu\nu}\gamma/2$,
with $\gamma=\eta^{\alpha\beta}\gamma_{\alpha\beta}$, then the linearized 
Einstein field equations yield
\begin{equation}
\bigcirc\gamma_{\mu\nu}=-2\kappa T_{\mu\nu},
\label{eq.41}
\end{equation}
along with the supplementary condition
\begin{equation}
\eta^{\rho\sigma}\gamma_{\mu\rho,\sigma}=0,
\label{eq.42}
\end{equation}
which solutions $\gamma_{\mu\nu}$ of Eq.~(\protect\ref{eq.41}) should satisfy. 
In Eq.~(\protect\ref{eq.41}) we
have used the notation for a generalized wave equation 
\begin{equation}
\bigcirc f=\eta^{\alpha\beta}f_{,\alpha\beta}=\left(\frac{1}{c^2}
\frac{\partial^2}{\partial t^2}-\nabla^2+\frac{1}{\tau^2}
\frac{\partial^2}{\partial v^2}\right)f.
\label{eq.43}
\end{equation}
Finally we see
from Eq.~(\protect\ref{eq.41}) that a necessary condition for 
Eq.~(\protect\ref{eq.42}) to be 
satisfied is 
that
\begin{equation}
\eta^{\alpha\beta}T_{\mu\alpha,\beta}=0,
\label{eq.44}
\end{equation}
which is an expression for the conservation of the energy and momentum without 
including gravitation.

The wave equation~(\protect\ref{eq.41}) is a generalization of the standard 
wave equation with
the D'Alambertian operator since, as seen from Eq.~(\protect\ref{eq.43}), it includes a 
second derivative with respect to the velocity. The velocity here is
actually the redshift parameter $z$. Thus gravitational waves depend on
spacetime and redshift (or velocity). This fact is not exhibited in Einstein's
general relativity theory.  

In vacuum, Eq.~(\protect\ref{eq.41}) reduces to
\begin{equation} 
\bigcirc\gamma_{\mu\nu}=0,
\label{eq.45}
\end{equation}
or
\begin{equation}
\left(\nabla^2-\frac{1}{c^2}\frac{\partial^2}{\partial t^2}\right)
\gamma_{\mu\nu}=\frac{1}{\tau^2}\frac{\partial^2\gamma_{\mu\nu}}{\partial v^2}.
\label{eq.46}
\end{equation}

Thus the gravitational field, like the electromagnetic field, propagates in
vacuum with the speed of light. The above analysis also shows the existance of
gravitational waves. 
 

\end{document}